\documentclass[fleqn,twoside]{article}
\usepackage{espcrc2}
\usepackage{graphicx}
\makeatletter
\newcommand\nameaddress[1]{{\addtocounter{address}\m@ne
                            \expandafter\xdef
                               \csname address.#1\endcsname
                               {\number\value{address}}%
                            \addtocounter{address}\@ne}}
\newcommand\useaddress[1]{{\edef\doit{\noexpand\addressmark
                                      \noexpand\setcounter{address}%
                                      {\number\value{address}}}%
                           \setcounter{address}{\csname address.#1\endcsname
                                                }%
                           \expandafter}\doit}
\newcommand\anotheraddress[1]{{\let\orig@makeadmark=\@makeadmark
                               \def\@makeadmark##1{\orig@makeadmark
                                                   {\negthinspace,##1}}%
                               \address{#1}}}
\newcommand\slashnext[1]{\mathpalette{\bgroup\let\style=}
                                     {\setbox0=\hbox{$\style #1$}%
                                      \setbox2=\hbox to\wd0{\hss$\style/$\hss}%
                                      \hbox to 0pt{\box2\hss}\box0\egroup}}
\@ifundefined{inlinecite}{}

\makeatother
\begin{document}
\title{\begin{flushright}\normalsize
            DUKE-TH-02-219 \\
       \end{flushright}
	Unexpected Results in the Chiral Limit with Staggered Fermions}
\author{Shailesh Chandrasekharan
\address{Department of Physics, Duke University, Durham, NC 27708-0305, USA}}

\begin{abstract}

A cluster algorithm is constructed and applied to study the chiral limit 
of the strongly coupled lattice Schwinger model involving staggered 
fermions. The algorithm is based on a novel loop representation of the 
model. Finite size scaling of the chiral susceptibility based on data from 
lattices of size up to $64\times 64$  indicates the absence of long range 
correlations at strong couplings. Assuming that there is no phase transition 
at a weaker coupling, the results imply that all mesons acquire a mass 
at non-zero lattice spacings. Although this does not violate any known 
physics, it is surprising since typically one expects a single pion to 
remain massless at non-zero lattice spacings in the staggered fermion 
formulation.

\end{abstract}
\maketitle

\section{Motivation}

Consider a two dimensional $U(1)$ lattice gauge theory with staggered 
fermions. This model has been studied extensively on the lattice
for over two decades \cite{Mar81,Car86,Dil95,Gut98,Gut99,Bie99,Far99}. 
In the limit when the lattice gauge coupling goes to zero, the model 
is believed to describe 
the continuum two flavor Schwinger model; the fermions are confined and 
the low energy physical particles are mesons. Further, in the chiral limit
three massless and one massive pseudo-scalar mesons emerge like in 
QCD \cite{Gat94}. However, since in two dimensions continuous chiral 
symmetries remain unbroken, the chiral condensate vanishes in the chiral 
limit. The actual prediction is
\begin{equation}
\langle \overline\psi\psi \rangle \;\sim\; m^{1/3},
\label{2fc}
\end{equation}
as discussed in \cite{Col76}. This means that the chiral susceptibility 
$\chi \;=\;\frac{\partial}{\partial m} \langle \overline\psi\psi \rangle$ 
diverges in the chiral limit reflecting the presence of massless 
excitations. This is consistent with the fact that $M_\pi \sim m^{2/3}$,
i.e., the pion mass vanishes at $m=0$. If one computes the 
susceptibility at $m=0$ in a finite box of size $L\times L$, one expects
\begin{equation}
\chi \;\sim\; L,
\label{ff}
\end{equation}
i.e., it diverges linearly with $L$ \cite{Smi96}.

When the lattice gauge coupling is non-zero, lattice artifacts 
in the staggered fermion formulation break the 
chiral symmetry of the two flavor Schwinger model explicitly down to a 
$U(1)$ subgroup. As far as we know, no one has completely analyzed the 
effects of lattice artifacts on the particle spectrum. Do massless particles 
remain in the spectrum? The conventional wisdom from higher dimensions
is that there should be one massless pion, since the remnant $U(1)$ 
chiral symmetry is expected to break spontaneously at strong couplings.
Of course the Mermin-Wagner theorem forbids the breaking of a continuous 
symmetry in two dimensions, although non-interacting massless bosons can 
emerge. On the other hand a $U(1)$ (or equivalently an $O(2)$) symmetry 
is special. In such a case 
bosons can interact through topological excitations as discovered by 
Kosterlitz and Thouless \cite{Kos73}. Thus, the lattice model could 
either be in a massive or a massless phase depending on the effective 
couplings of the low energy effective theory describing the bosonic 
excitations. In the massless phase there are predictions for the 
behavior of the chiral condensate and the susceptibility
based on universality \cite{Zinn}. One expects
\begin{equation}
\langle \overline\psi\psi \rangle \;\sim\; m^{\eta/(4-\eta)}.
\label{ktc}
\end{equation}
where $0 \leq \eta \leq 0.25$. Again the susceptibility $\chi$ diverges
in the chiral limit reflecting the presence of massless excitations.
In this case the finite size scaling formula for the susceptibility 
is given by
\begin{equation}
\chi \;\sim\; L^{2-\eta}.
\label{kt}
\end{equation}
It is amusing that if we set $\eta=1$ in eqs. (\ref{ktc}) and 
(\ref{kt}) we recover eqs. (\ref{2fc}) and (\ref{ff}).

Most results from earlier work on the lattice Schwinger model favor 
the view point that there is one massless pion at finite lattice 
spacings. This is based on the observation that the pion mass 
decreases with the fermion mass like in the continuum Schwinger 
model. However, on closer examination the lattice model
shows deviations, which become larger at stronger couplings as 
one might expect \cite{Gut98,Gut99}. On the 
other hand, no one has ever found scaling relations expected 
from universality in a massless phase of an $O(2)$ model in two 
dimensions. In particular no one has been able to confirm relations 
similar to eqs. (\ref{ktc}) or  (\ref{kt}). Thus, inspite of the 
large amount of literature on the subject, questions related to 
the chiral limit of the lattice Schwinger model with staggered 
fermions still remain unanswered.

The essential difficulty is the absence of a reliable numerical 
approach to study interacting fermionic field theories close to the 
chiral limit. For this reason most previous studies alluded to above, 
have focused on calculations away from the chiral limit and have used 
extrapolation techniques to predict the chiral limit. As we will see
this can be misleading. In the last few years fermion algorithms based
on cluster updates have emerged, which do not suffer from problems that 
conventional algorithms face near the chiral limit \cite{Cha99}. One 
can now work directly in the chiral limit using the new approach, a 
luxury not available with earlier methods. Recently, a fermion
cluster algorithm has confirmed the scaling predictions near a 
Kosterlitz-Thouless transition in a Hubbard type model with unmatched 
precision \cite{Cha01.a}. Interestingly, these new ideas can also be 
applied to study the chiral limit of the lattice Schwinger model at 
strong couplings \cite{Cha01}. The algorithm is based on a novel 
loop representation of the model. Although, it may be possible to 
extend the method to weaker couplings, in this article we concentrate 
on the strong coupling limit. We focus on the question of whether there 
are massless excitations at strong couplings by looking for a divergence
in the chiral susceptibility as predicted by eq.(\ref{kt}).

\section{The Model}

The two dimensional $U(1)$ lattice gauge theory with staggered fermions
is described by the action
\begin{eqnarray}
S &=& -\frac{1}{g^2} \sum_{x,\mu,\nu,\mu\neq \nu}  
\mathrm{Re}
[U_\mu(x)U_\nu(x+\hat{\mu})U^\dagger_\mu(x+\hat{\nu})U^\dagger_\nu(x)]
\nonumber \\
&& + \sum_{x,\mu}\;\overline{\psi}_x \eta_\mu(x) \left(
U_\mu(x) \psi_{x+\hat{\mu}} - U^\dagger_\mu(x-\hat{\mu}) \psi_{x-\hat{\mu}}
\right)
\nonumber \\
&& \hskip0.5in +\;m \sum_x \overline{\psi}_x \psi_x,
\end{eqnarray}
where $x\equiv (x_1,x_2)$ represents a lattice site on an 
$L\times L$ lattice, $\mu,\nu=1,2$ represent the two 
directions and $\hat{\mu},\hat{\nu}$ are the unit vector in the 
$\mu$ and the $\nu$ direction respectively. The phase factors 
$\eta_1(x) = 1$ and $\eta_2(x)= (-1)^{x_1}$ are the staggered 
fermion phase factors. The gauge fields $U_\mu(x)$ are elements of 
$U(1)$ phase factors and the fermion fields $\psi(x)$ and 
$\overline{\psi}(x)$ are Grassmann numbers.

In the strong coupling limit ($g=\infty$) it 
is possible to integrate over the gauge fields and obtain
\begin{eqnarray}
Z(m) &=&\;\int [d\psi][d\overline\psi]\; \prod_x\; 
\left(1 - m \overline{\psi}_x \psi_x\right)
\nonumber \\
&& \;\;\;\;\;\;\;\;\;
\prod_{x,\mu}\;
\left(1 + \overline{\psi}_x \psi_x \overline{\psi}_{x+\mu} \psi_{x+\mu}\right),
\end{eqnarray}
which can be written as a sum over weights of configurations of monomers 
and dimers \cite{Ros84}. Each configuration consists of $n_x=0,1$ monomers 
on site $x$ and $b_{x,\mu} = 0,1$ dimers on the bond connecting $x$ and 
$x+\hat{\mu}$. In order for the Grassmann integration to give a non-zero 
result we need
\begin{equation}
n_x + b_{x,1} + b_{x,2} + b_{x-\hat{1},1} + b_{x-\hat{2},2}\;=\; 1.
\label{constraint}
\end{equation}
\begin{figure}[ht]
\begin{center}
\includegraphics[width=0.4\textwidth]{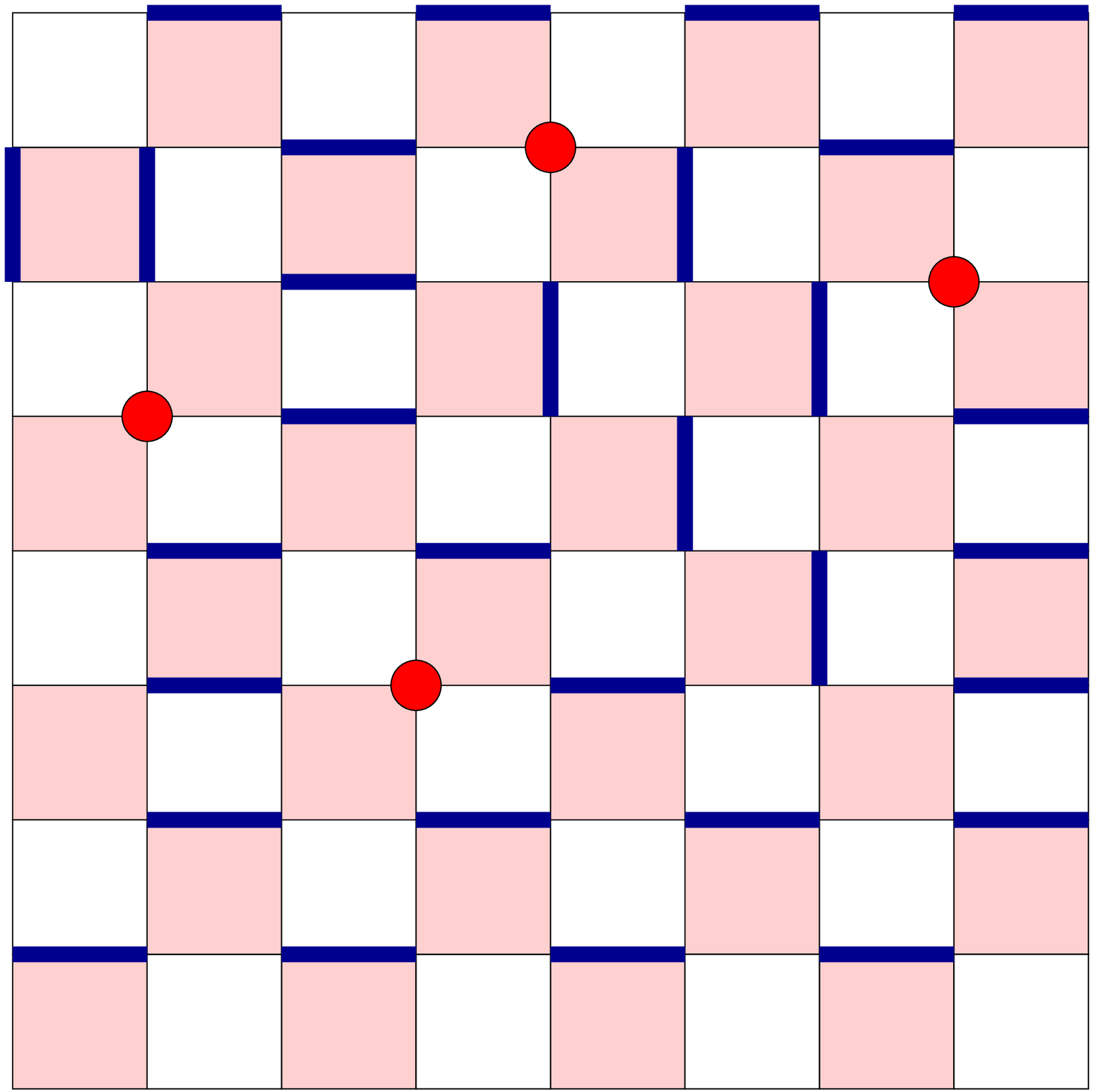}
\end{center}
\caption{{\em \label{mconf} A typical monomer-dimer configuration 
satisfying eq.(\ref{constraint}). The weight of the configuration is 
$m^4$.}}
\end{figure}
Assuming this constraint the partition function can be written as
\begin{equation}
Z(m) \;=\; \sum_{[n,b]} \exp\left[ \log(m) \; \sum_x \;n_x\right]
\end{equation}
A typical configuration is shown in figure \ref{mconf}. 

When $m=0$ no monomers are allowed and the partition function is given 
by the number of closely-packed-dimer (CPD) configurations on the 
lattice. Such configurations are interesting even in condensed matter 
physics and have played an important role in the study of the 2-d Ising 
model \cite{Mc73}. The chiral symmetry of staggered fermions is manifest 
at finite volumes through the fact that the chiral condensate vanishes 
since it is impossible to find a CPD configuration with just one monomer. 
The chiral susceptibility on the other hand is non-zero and is a useful
observable in the chiral limit. It is equal to the total number of CPD 
configurations with two monomers, where one of the monomers is 
constrained to be at a fixed position, divided by the partition function. 

\section{Loop Representation}

It is easy to construct a local Metropolis algorithm for the monomer-dimer
model when $m\neq0$. The algorithm is based on an update which
either breaks a dimer into two monomers or vice-versa. This algorithm
works reasonably well for $m \geq 0.01$. However, the algorithm 
fails in the chiral limit since no monomers are allowed when $m=0$.
In fact in the chiral limit it is easy to find configurations where 
simple local updates do not lead to another allowed configuration. Perhaps
for this reason, as far as we know, no one has successfully studied the 
chiral limit.

\begin{figure}[hbt]
\begin{center}
\includegraphics[width=18pc]{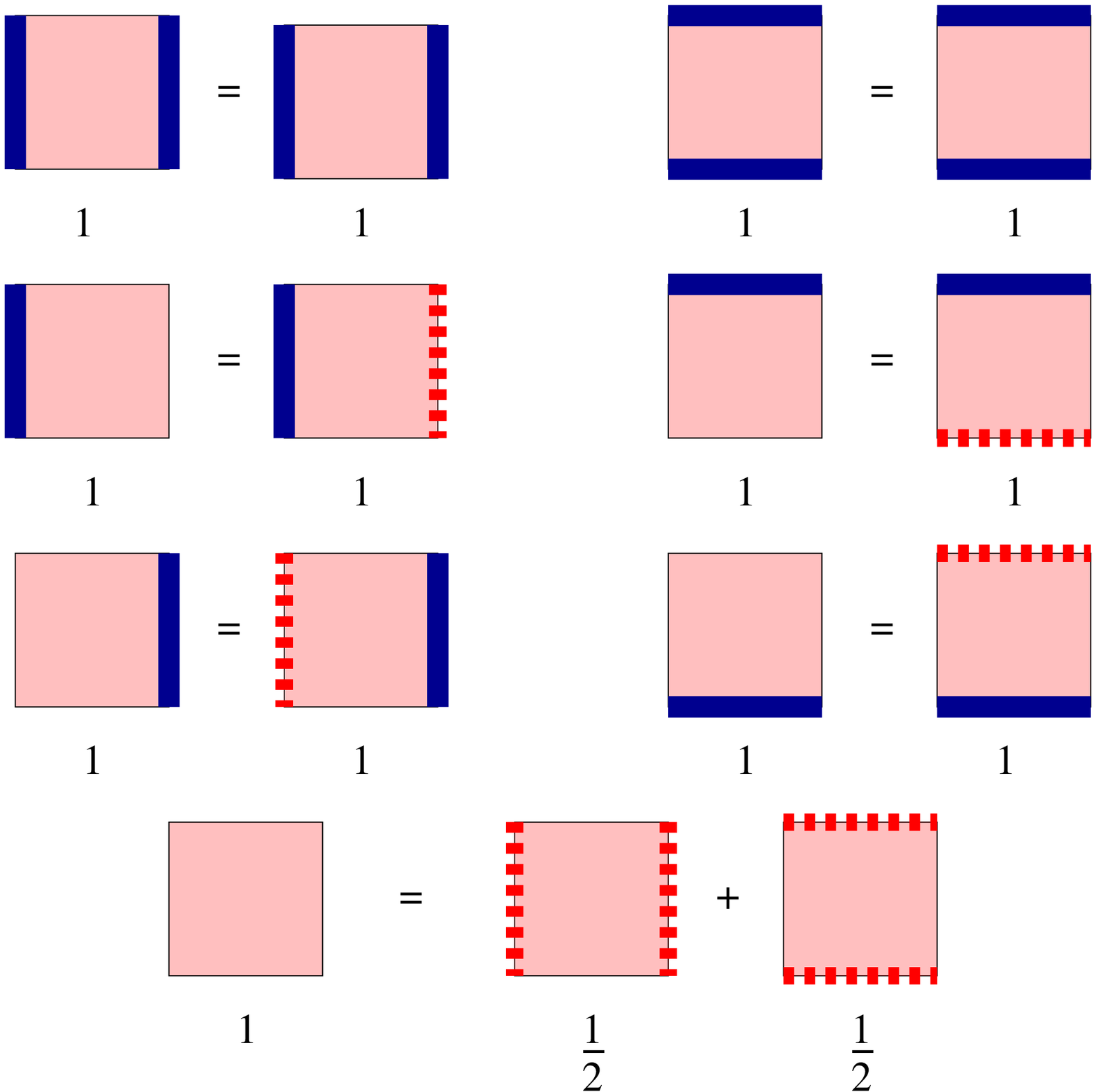}
\caption{ \label{rules} \em Rules for extending the CPD configurations to
include additional bond variables.}
\end{center}
\vskip-0.3in
\end{figure}

\begin{figure}[htb]
\begin{center}
\includegraphics[width=18pc]{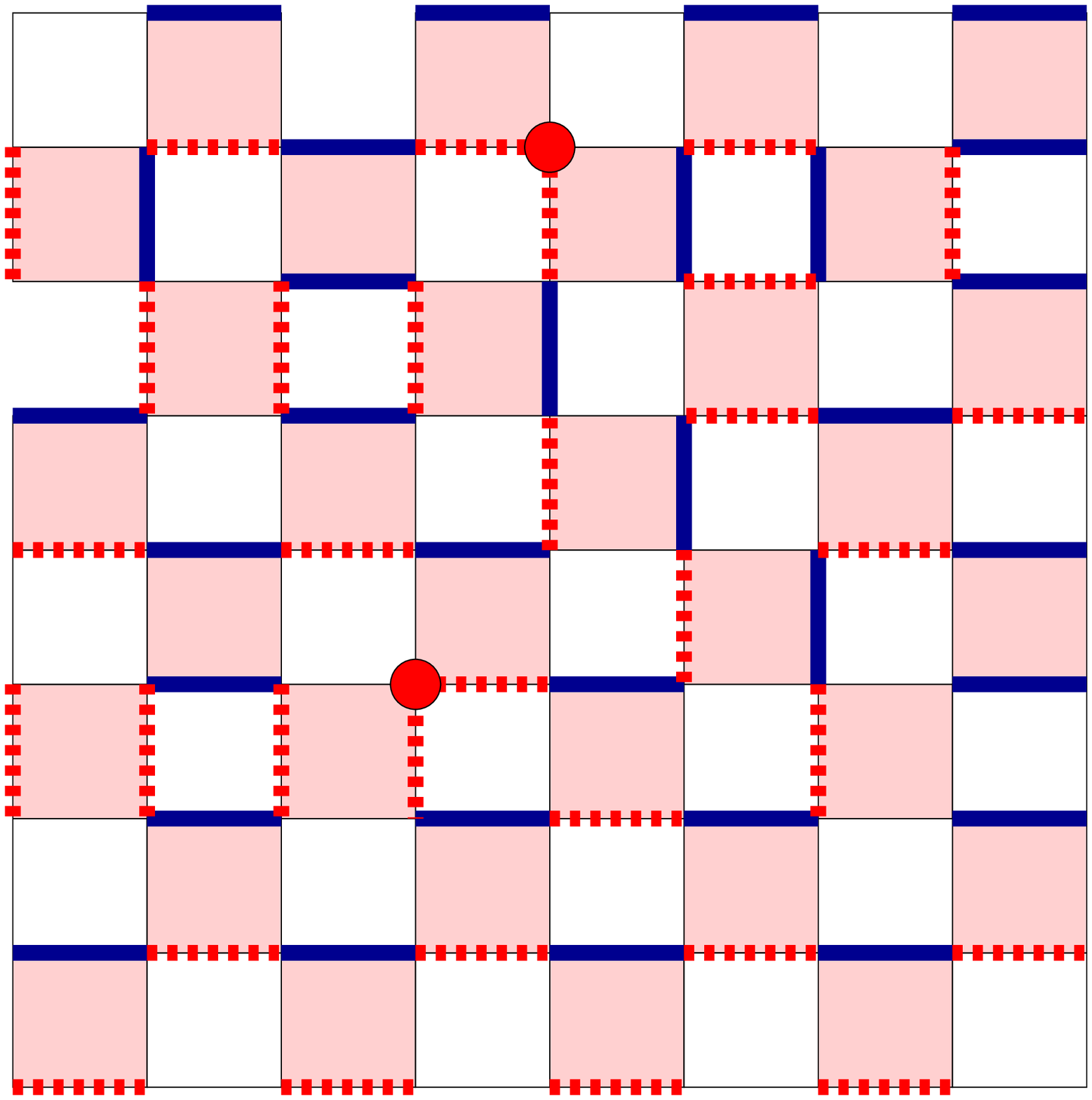}
\caption{ \label{loopconf} \em An example of a loop configuration 
with two monomers.}
\end{center}
\vskip-0.4in
\end{figure}

To construct an algorithm that is applicable in the chiral limit 
we extend CPD configurations to configurations of loops made up of bonds
which include the original or ``filled'' dimers (represented here by 
``solid'' bonds) and ``empty'' dimers (represented by ``dashed'' bonds) 
such that the partition function can be expressed as a sum over weights 
of new loop configurations. Figure \ref{rules} shows the rules of one 
such extension. If we ignore monomers each shaded 
plaquette of the configuration of Fig. \ref{mconf} carries one of the 
seven plaquette configurations given on the left side of equations in 
Fig. \ref{rules}. On the other hand, the right side of the equations 
represent the new configurations and their weights. Figure \ref{loopconf} 
gives an example of a loop 
configuration. 
In the absence of monomers it is easy to check that
all constraints are satisfied if each loop is made up of a repeating 
sequence of filled and empty dimers. This means that there are two
allowed configurations associated with each loop. The usefulness of the 
loop variable is that a dimer system can be updated by ``flipping'' a 
loop where filled dimers are emptied and vice versa. The Metropolis 
acceptance of such a loop flip is found to be reasonable although large
loops are not flipped as often as small loops.
When monomers are allowed each loop consists of an even number of 
them. Further, when monomers are present in a loop then it contains
a unique pattern of filled and empty dimers and a flip is not allowed. 
However, close to the chiral limit loops containing monomers are
negligible. 

The algorithm to measure the susceptibility in the chiral limit is 
straight forward in the loop representation. Typically, we choose 
a point at random on the lattice and traverse the loop it is on,
starting in the direction of a filled bond. As one goes around the 
loop, the bonds are flipped and the change in the weight of the 
configuration is noted. Interestingly, every time a filled bond is 
emptied one gets a configuration that contributes to the susceptibility. 
Since its weight relative to the original configuration is known 
at the time of the flip, it is recorded as a part of the measurement. 
When the whole loop is flipped one knows the change in the weight of 
the configuration and a Metropolis accept-reject step can be performed. 
If the configuration is accepted then one gets a new configuration. 
Otherwise the loop is flipped back to the original configuration. 
In any case, the sum over all the weight changes recorded while flipping 
the filled bonds in the loop divided by one plus the total weight 
change due to the loop flip is taken as a contribution to the 
susceptibility during that update. It is possible to show that this
algorithm is ergodic.

\section{Results}

Since the remnant $U(1)$ chiral symmetry of staggered fermions
can break spontaneously
in higher dimensions, the chiral condensate can approach 
a constant in the chiral limit. However, in two dimensions a continuous 
symmetry cannot break and it is almost guaranteed that
\begin{equation}
\langle \overline\psi\psi \rangle \;\sim\; m^\delta.
\label{ccpow}
\end{equation}
with $0 < \delta \leq 1$. Comparing with eqs. (\ref{2fc}) and 
(\ref{ktc}) we see that $\delta = 1/3$ in the continuum 
two flavor Schwinger model and $\delta \leq 1/15$ is expected from 
$O(2)$ universality in two dimensions. It is important to remember 
that the $U(1)$ chiral symmetry of staggered fermions is non-anomalous. 
Thus, a non-zero expectation value of $\langle \overline\psi\psi \rangle$
in the chiral limit would definitely imply a spontaneous breaking of
the $U(1)$ symmetry.

\begin{figure}[htb]
\vskip0.1in
\begin{center}
\includegraphics[width=20pc]{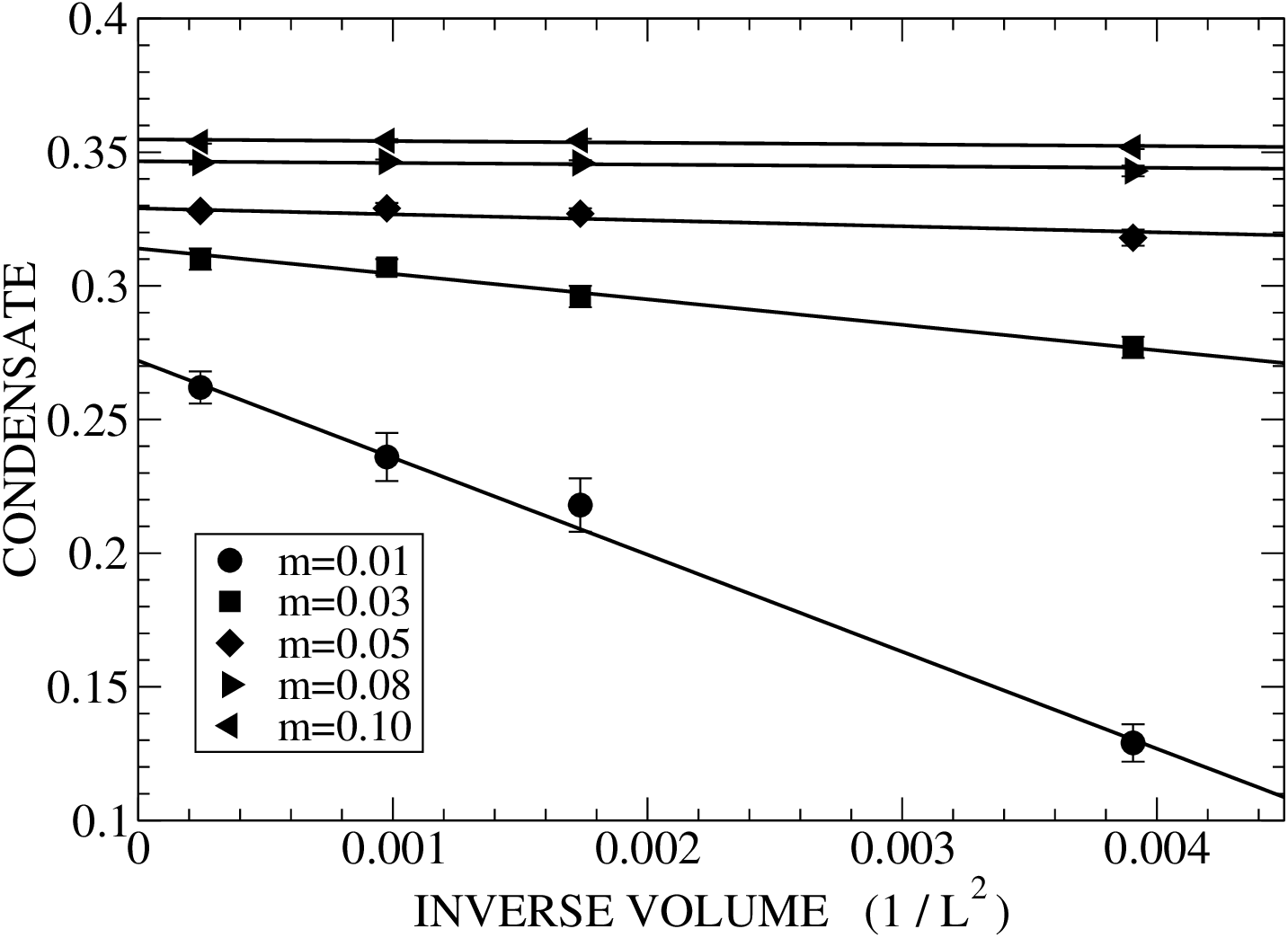}
\caption{ \label{fspsi} \em Chiral Condensate as a function of inverse 
volume at various masses.}
\end{center}
\vskip-0.3in
\end{figure}

\begin{figure}[htb]
\vskip0.1in
\begin{center}
\includegraphics[width=20pc]{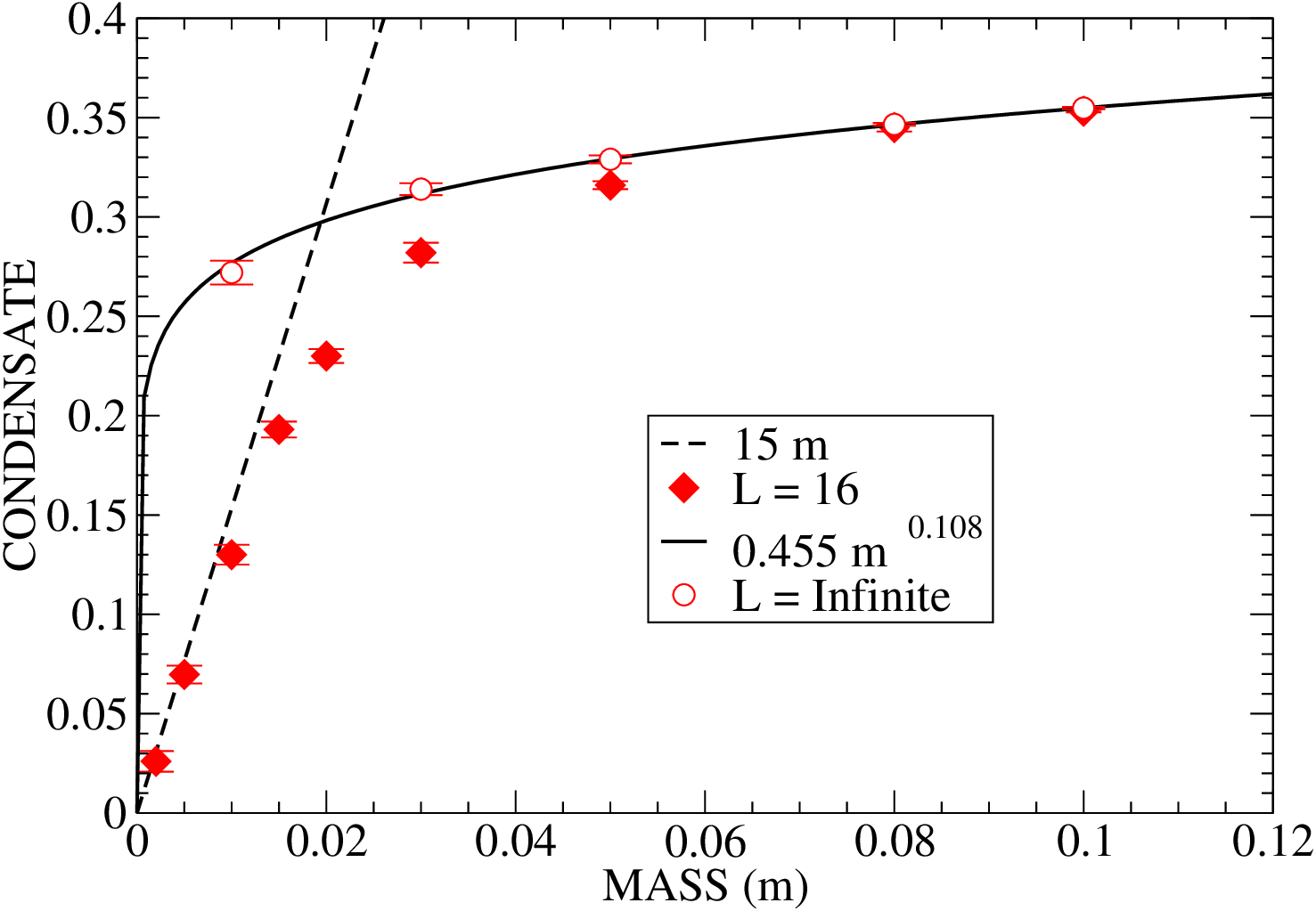}
\caption{ \label{psi} \em Chiral condensate as a function of mass.}
\end{center}
\vskip-0.3in
\end{figure}

Using the local Metropolis algorithm we calculated the chiral condensate for 
five masses in the region $0.01 \leq m \leq 0.1$ for $L=16,24,32$ and
$64$.
Below $m = 0.01$ the algorithm slows down and is not very efficient.
Our results are shown in figure \ref{fspsi}. The solid lines are linear
fits of the data to the finite size scaling formula 
$\langle \overline\psi\psi\rangle = A + B/L^2$ at a fixed value of $m$. 
The constant $A$ then yields the condensate at infinite volume for a
fixed mass. 
Figure \ref{psi} shows the condensate as a function of the mass
for $L=16$ and for infinite $L$ obtained from fits
shown in figure \ref{fspsi}. At $L=16$ we see that the condensate goes 
linearly to zero as expected due to finite volume effects\footnote{ 
In a finite box the partition function $Z(m)$ is a polynomial of
the variable $m^2$. The condensate is proportional to the first
derivative of $Z(m)$ with respect to $m$ and hence vanishes linearly 
with $m$.}.
On the other hand the infinite volume results fit beautifully 
to a power law of
the form $\langle \overline\psi\psi\rangle = 0.455(6)m^{0.108(5)}$.
Since we are in the strong coupling limit it is not surprising that the 
power does not match the continuum two flavor result of eq. (\ref{2fc}).
However, it neither matches the predictions of the two 
dimensional $O(2)$ model based on universality (see eq.(\ref{ktc})).

\begin{figure}[htb]
\vskip0.1in
\begin{center}
\includegraphics[width=20pc]{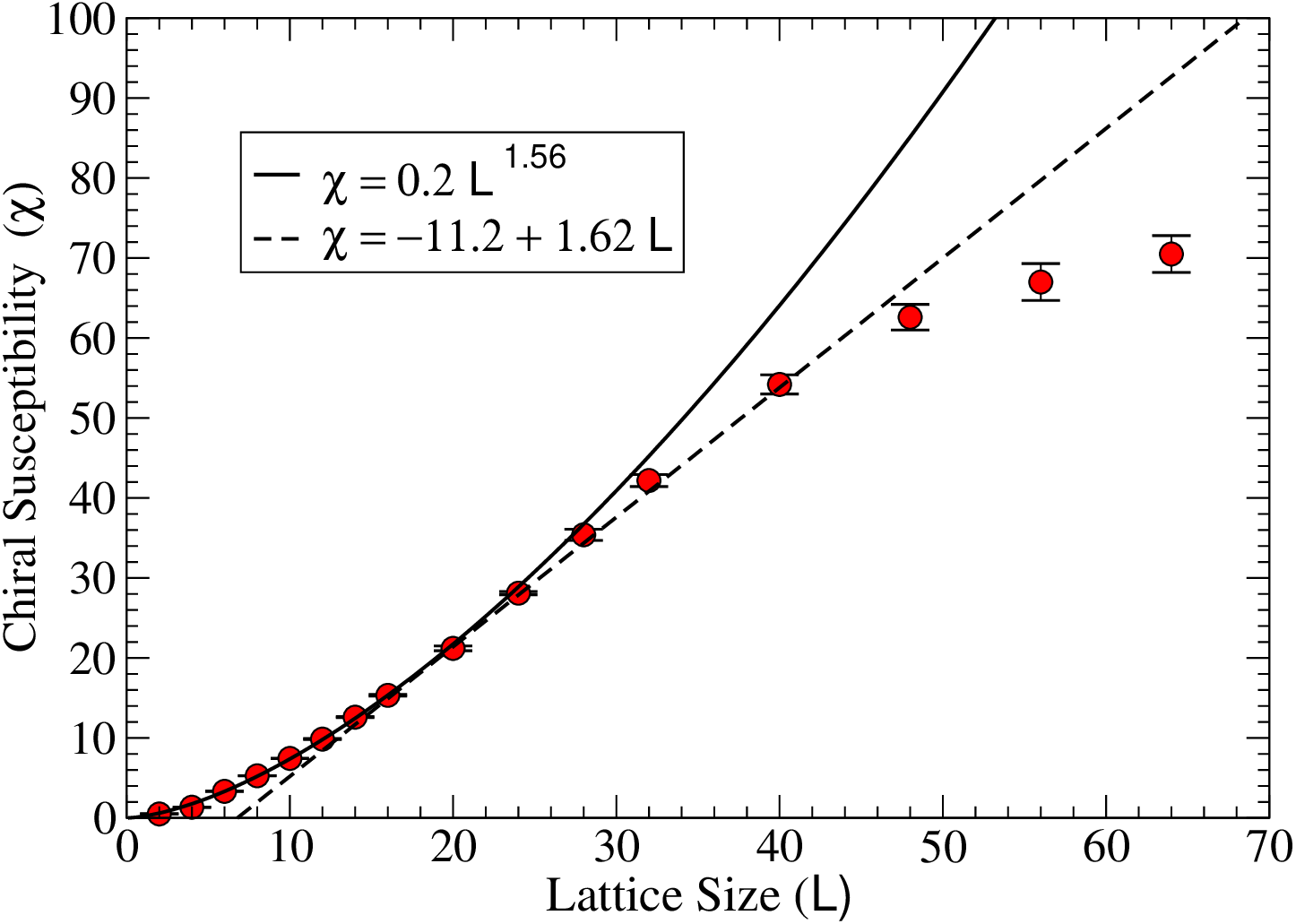}
\caption{ \label{sus} \em Finite size scaling of the chiral susceptibility.}
\end{center}
\vskip-0.3in
\end{figure}

Before attempting to understand the unexpected value of the power
it is useful to confirm that the power law behavior is valid all
the way to the chiral limit. A power law for the chiral condensate 
implies that the chiral susceptibility will diverge at $m=0$ in 
the thermodynamic limit. This divergence typically manifests itself 
in a finite size scaling of the form
\begin{equation}
\chi = \frac{\partial}{\partial m} \langle \overline\psi\psi \rangle 
\;\sim\; L^\gamma,
\label{csus}
\end{equation}
consistent with eqs. (\ref{ff}) and (\ref{kt}). This behavior
is difficult to study with conventional algorithms. However, the 
loop cluster algorithm discussed in the earlier section
is designed exactly for this purpose. In figure \ref{sus} we plot 
our results for the chiral susceptibility obtained with the new
algorithm at $m=0$ as a function of the lattice size.
Notice that for $L=16$ the value of the susceptibility is consistent 
with $15$, the slope of the condensate at $m=0$ shown in figure \ref{psi}.

As a function of $L$ the susceptibility behaves differently in
different regions. The data is described well by a power law in the 
region $8\leq L \leq 20$ with $\gamma=1.56$, while a straight line 
fits the data in the region $20\leq L \leq 40$. For $L > 40$ 
the susceptibility appears to be approaching a constant. If we 
compare the power law behavior of the susceptibility in the region 
$8\leq L \leq 20$ with eq. (\ref{kt}) we find that $\eta \sim 0.44$ 
which is inconsistent with the predictions of $O(2)$ universality
in which one would expect $0\leq \eta \leq 0.25$. However, it explains 
the strange power law behavior of the infinite volume chiral condensate 
we encountered in figure \ref{psi}. If we substitute $\eta=0.44$ in 
eq. (\ref{ktc}) 
we find that $\langle \overline\psi \psi\rangle \sim m^{0.12}$, 
which is quite close to the $L=\infty$ results of figure \ref{psi} 
in the region $0.01\leq m \leq 0.1$. Since the power law behavior
of the susceptibility does not extend to larger volumes, it is
clear that the power law behavior of the infinite volume chiral 
condensate shown in figure \ref{psi} cannot hold all the way to 
the chiral limit. In fact the susceptibility at $m=0$ saturates 
at large volumes. Equivalently, the first derivative of the chiral 
condensate with respect to the mass at $m=0$ reaches a constant as 
the volume becomes large\footnote{Although we cannot rule out a mild
divergence of the susceptibility in figure \ref{sus}, we think 
it is unlikely.}. This means the infinite volume chiral condensate 
will also vanish linearly
at $m=0$. Typically, divergence of a susceptibility such as the one
defined in eq.~(\ref{csus}), is related to the presence of massless 
particles in the theory.  Our results indicate that there are no
such particles.

\section{Discussion}

The results that we have obtained are somewhat surprising. The 
conventional wisdom is that strong couplings break chiral symmetry 
at least in three or more dimensions. This suggests that the low energy 
effective model that describes the $U(1)$ chiral dynamics of the 
lattice Schwinger model with staggered fermions is most likely in the 
low temperature (massless) phase at strong couplings. This reasoning 
appears to be in contradiction with our findings since we do not find 
the expected divergence described by eq. (\ref{kt}) which in turn 
implies that there are no massless particles at strong couplings.

Is it possible we have missed something? Of course, we rely heavily on
our algorithm. It is rather new and may have weaknesses like long 
auto-correlation times that we have not yet appreciated. In order to 
alleviate such fears we have compared our Monte Carlo results with 
exact calculations of the susceptibility at $m=0$ on small lattices. 
Table \ref{exact} shows this comparison. 
\begin{table}[ht]
\vskip-0.1in
\begin{center}
\caption{ {\label{exact} Chiral Susceptibility: algorithm vs. exact results.}}
\vskip0.1in
\begin{tabular}{|c|c|c|}
\hline
Lattice Size & Exact & Algorithm \\ 
\hline
$4 \times 4$ & 1.70588235... & 1.7059(1) \\
$6 \times 6$ & 3.33640880... & 3.3364(1) \\
$6 \times 8$ & 4.07961565... & 4.0796(2) \\
$8 \times 8$ & 5.27221660... & 5.2722(2) \\
\hline
\end{tabular}
\end{center}
\vskip-0.3in
\end{table}
Clearly, we can reproduce the exact results with great precision at least
on small lattices. We have not seen any pathologies in the simulations 
at larger volumes except for the fact that the fluctuations increase which
require us to increase the statistics proportionally. Based on this we
are prejudiced to believe in our estimate of the errors within a factor 
of two or three.

Assuming our results are correct, we find that at strong couplings
the lattice model is in the high temperature (massive) phase of the low
energy effective model. We know from experience in higher dimensions 
that chiral symmetry breaking effects become weaker at smaller couplings.
This, coupled with the fact that there is no deconfinement transition in 
two dimensions, suggests that it is unlikely that there is a phase
transition to a massless phase at weaker couplings. Our results then 
imply that there are no massless excitations 
for all $g\neq 0$ (or equivalently at finite lattice spacings); the 
non-singlet pseudo-scalar meson (which would be expected to become 
the Goldstone boson in higher dimensions ) is actually massive at all
non-zero couplings. This conclusion may come as a surprise to some, 
but it does not contradict any known physics as far as we can tell. 
In particular it does not contradict our expectations that in the 
zero coupling limit one must recover the continuum two flavor Schwinger 
model. It is well known that two other non-singlet pseudo-scalar mesons 
acquire a mass due to lattice artifacts in the staggered fermion 
formulation. Our data is suggesting that all mesons become 
massive at non-zero couplings contrary to expectations from higher 
dimensions where the Goldstone boson will remain massless at finite 
lattice spacings. In the lattice Schwinger model this boson is still 
perhaps the lightest. The gauge coupling controls the lattice spacing 
in the model; typically one uses the mass of the iso-singlet meson to 
set the lattice spacing. This means that in order 
to recover the correct continuum limit the singlet and the non-singlet 
meson masses must become smaller in lattice units as one approaches 
weaker couplings but their ratio must diverge. 

The current work can be extended in many directions. There are results
in the large $N_c$ and large $d$ limits at strong couplings which show 
that chiral symmetry is indeed broken \cite{Kaw81,Klu81}. Here we have 
shown that at $N_c=1$ and $d=2$ chiral symmetry remains unbroken and 
the theory is in the massive phase. It would be interesting to find the
value of $N_c$ for $d=2$ at which the theory will move into a phase 
with massless excitations. There is a lot of evidence from lattice 
simulations and chiral extrapolations
that shows lattice QCD with staggered fermions to be in the chirally
broken phase. However, as we have seen in the present 
context, chiral extrapolations can sometimes be misleading. Fortunately, 
cluster algorithms of the type used here can be developed for any $N_c$ 
and $d$ so that these questions can be answered directly in the chiral
limit. One can start with the results presented in \cite{Kar89}.
It would be useful to confirm that chiral symmetry is indeed
broken at $N_c=3$ and $d=4$. On the algorithmic side, it is exciting 
that we have a new method to explore the chiral limit in certain gauge 
theories. Loop variables play an important role in this method. Such 
variables have already been discovered in a limited class of 
models \cite{Sal91,Chr99}. Perhaps it is possible to extend them to 
other interesting models.

\section*{Acknowledgment}

I would like to thank C. Gattringer, C.B. Lang, 
J.C. Osborn, J.J.M. Verbaarschot and U.-J. Wiese for helpful 
comments. This work was supported in part by US Department of 
Energy funds under the grant DE-FG02-96ER40945. The computations 
were performed on BRAHMA, a Pentium based Beowulf cluster 
funded by grants from the Intel Corporation.

\end{document}